\documentclass[aps,pra,twocolumn,superscriptaddress,showpacs,floatfix]{revtex4-1}
\usepackage{graphicx,amsmath,amssymb,xspace,epsfig,float,multirow}

\begin{document}

\title{Exact solution for the degenerate ground-state manifold of a strongly interacting one-dimensional Bose-Fermi mixture}

\author{Bess Fang}
\affiliation{Department of Physics, Block S12, Faculty of Science, National
  University of Singapore, 2 Science Drive 3, Singapore 117542}
\affiliation{Centre for Quantum Technologies, National University of
  Singapore, 3 Science Drive 2, Singapore 117543}

\author{Patrizia Vignolo}
\affiliation{Universit\'e de Nice-Sophia
  Antipolis, Institut Non Lin\'eaire de Nice, CNRS; 1361 route des Lucioles, 06560 Valbonne, France}
\author{Mario Gattobigio}
\affiliation{Universit\'e de Nice-Sophia
  Antipolis, Institut Non Lin\'eaire de Nice, CNRS; 1361 route des Lucioles, 06560 Valbonne, France}
\author{Christian Miniatura}
\affiliation{Department of Physics, Block S12, Faculty of Science, National
  University of Singapore, 2 Science Drive 3, Singapore 117542}
\affiliation{Centre for Quantum Technologies, National University of
  Singapore, 3 Science Drive 2, Singapore 117543}
\affiliation{Universit\'e de Nice-Sophia
  Antipolis, Institut Non Lin\'eaire de Nice, CNRS; 1361 route des Lucioles, 06560 Valbonne, France}
\author{Anna Minguzzi}
\affiliation{Universit\'e Grenoble 1, CNRS, LPMMC, UMR5493, Maison des Magist\`{e}res, 38042 Grenoble,
  France}

\date{\today}

\begin{abstract}
We present the exact solution for the many-body wavefunction of a
one-dimensional mixture of bosons and spin-polarized fermions with equal
masses and infinitely strong repulsive interactions under external
confinement. Such a model displays a large degeneracy of the ground
state. Using a generalized Bose-Fermi mapping we find the solution for the
whole set of ground-state wavefunctions of the degenerate manifold and we
characterize them according to group-symmetry considerations. We find that the
density profile and the momentum distribution depends on the symmetry of the
solution.  By combining the wavefunctions of the degenerate manifold with suitable symmetry and guided by the strong-coupling form of the Bethe-Ansatz solution for the homogeneous system we propose an analytic expression for the many-body wavefunction of the inhomogeneous system which well describes the ground state at finite, large and equal interactions strengths, as validated by numerical simulations. 
\end{abstract}


\maketitle

\section{Introduction} 
Ultracold atomic gases provide a versatile and controlled system for the study
of quantum correlations and fluctuations which are particularly strong in one
dimension (1D). Experiments on two-dimensional optical lattices
\cite{Paredes04,Kinoshita04,Kinoshita06,Palzer09} or on a chip trap
\cite{VanDruten08} have reached the strongly interacting Tonks-Girardeau
regime.  In such impenetrable boson limit, repulsive interactions play the
role of the Pauli exclusion principle and the many-body wavefunction can be
exactly obtained by mapping onto the one of noninteracting fermions
\cite{Gir1960}. Experimental advances on trapping and cooling ultracold
Bose-Fermi mixtures
\cite{Schreck01,Truscott01,Hadzibabic02,Silber05,Ospelkaus06,Fukuhura09} and
the possibility of trapping both species in tight atomic waveguides have
boosted a theoretical activity on 1D mixtures. At increasing boson-fermion
repulsions, mean-field \cite{Das03} and Luttinger liquid analysis at weak
coupling \cite{CazHo03} predict an instability towards phase separation
(i.e. demixing) of the two components.  For a highly symmetric model with
equal masses and coupling constants, further progress can be made by use of
exact solutions.  For the homogeneous system, a Bethe-Ansatz solution is known
\cite{LaiYang,Batchelor05,ImaDem06} and no demixing is found.  The
long-wavelength properties of its correlation functions have been studied
using conformal field theory \cite{FraPal05}. Inhomogeneous systems, as in the
case of experiments, bring about novel issues, such as the spatial structure
of the ground state. At intermediate  interaction strength a partial demixing of the
two clouds has been found by a local density approximation on the Bethe-Ansatz
solution both at zero and finite temperature \cite{ImaDem06,ImaDemAnn,Yin09}.
In the Tonks-Girardeau limit of infinitely strong boson-boson and
boson-fermion repulsions a large ground state degeneracy is expected
\cite{GirMin07}, and is associated to the freedom of fixing the sign of the
many-body wavefunction under the exchange of a boson with a fermion. For an
inhomogeneous system, one exact solution of the degenerate manifold has been
proposed in \cite{GirMin07} and analyzed in detail in
\cite{Fang09,Lelas09,Lu10}.  The corresponding density profiles display no
demixing among the two species.  Till now no expression 
was known for the other wavefunctions of the manifold.
 In this
work we solve several open theoretical issues. First of all we find an exact
analytical solution for all the wavefunctions of the degenerate manifold in
the Tonks-Girardeau limit, thus generalizing the solution of
\cite{GirMin07}. Secondly, we characterize the solutions in terms of their
symmetry properties according to group theory considerations. At difference from 
fermionic or bosonic spinor systems \cite{Deuretzbacher,Guan09}, where the
state of the system can be labelled on the basis of the spin quantum number,
 in order to label the states of the
Bose-Fermi mixture we  introduce a suitable Casimir operator which reflects the mixed
symmetry under particle exchange. Furthermore, we find that such symmetry considerations
allow for the understanding of the shape of the momentum distribution, which
depends on the choice of the wavefunction within the manifold. Finally, by linear combination of the basis wavefunctions of  the degenerate manifold  we individuate the wavefunction which corresponds to the ground state at finite, large and equal interactions strengths, and we confirm this prediction by comparing with numerical DMRG simulations.
For the nontrivial case of the Bose-Fermi mixture with large degeneracy, this
analysis allows for the first time  to draw a link   between the
Bethe-Ansatz solution of the homogeneous system and the Tonks-Girardeau
solution of the inhomogeneous system. Our solution sheds light onto the
general properties of the ground state wavefunction of a fully quantum problem
in the strongly interacting limit.

\section{
Orthonormal basis set for the degenerate ground-state manifold} 
\subsection{General considerations}

We consider the model of $N_B$ bosons and $N_F$ spin-polarized fermions of
masses $m_B=m_F=m$, confined by the same external potential.
 The particles interact via the contact potentials,
$v_{BB}(x)=g_{BB}\delta(x)$, $v_{BF}(x)=g_{BF}\delta(x)$, and we focus on the
limit $g_{BB}=g_{BF}\to \infty$.  The effect of contact interactions can be
replaced by the boundary condition that the wavefunction vanishes at each
BB or BF contact, i.e. 
\begin{equation}
\Psi(..,x_j,..,x_\ell,..)\!=\!0~\mathrm{whenever}~x_j\!=\!x_\ell.  
\end{equation}
 We adopt the convention that $\{x_1,...,x_{N_B}\}$ are bosonic
coordinates and  $\{x_{N_B+1},...,x_{N}\}$ are fermionic ones. 
The ground state has a large degeneracy $C^N_{N_B}=N!/N_B!/N_F!$, which can be
interpreted as choosing $N_B$ positions for the bosons out of $N=N_B+N_F$, and 
amounts to fixing in several possible ways the sign of the
wavefunction under the exchange of bosons with fermions.

In order to determine an orthonormal basis set for the degenerate manifold we
proceed as follows. Consider a fermionic Slater determinant made of the first
{\em total} $N$ orbitals,
\begin{equation}\label{eqn:SD}
\Psi_F(x_1,...,x_N)=\frac{1}{\sqrt{N!}} \det[\phi_j(x_\ell)],
\end{equation} 
where $j,\ell=1,...,N$, and $\phi_j(x)$ are obtained by the
solution of the single-particle Schroedinger equation in the potential
$V(x)$. $\Psi_F$ displays the correct nodes at each $BB$ and $BF$ contact.  In
a given coordinate sector, $x_{P(1)}<x_{P(2)}<...<x_{P(N)}$, with $P$ being a
permutation among the $N$ particles, the required many-body wavefunction is
proportional to $\Psi_F(x_1,...,x_N)$. A useful set of orthonormal basis is
given by the ``snippets'' \cite{Deuretzbacher}
\begin{equation}
\langle x_1, ..., x_N|P\rangle=\sqrt{N!}|\Psi_F(x_1,...,x_N)|
\end{equation} 
in the coordinate sector $x_{P(1)}<x_{P(2)}<...<x_{P(N)}$, and zero
otherwise. To build the required basis for the manifold, we now collect
the snippets which correspond to exchanging only the positions of the
bosons or of the fermions {\em among themselves}, using the Fermi or Bose
statistics to fix the relative sign of the various terms. The number of groups
of snippets subdivided in such a way is exactly 
$C^N_{N_B}$.  This yields the
required orthonormal basis set $\{ \Psi^\alpha \}$ for the degenerate manifold, 
since each snippet is orthogonal to another and is used only once.

The density profiles associated to each wavefunction $\Psi_\alpha$ are 
given by 
\begin{eqnarray}
n_B^\alpha(x) & = & N_B \int dx_2...dx_N |\Psi_{\alpha}(x,x_2..x_N)|^2,\nonumber\\
n_F^\alpha(x) & = & N_F \int dx_1...dx_{N-1} |\Psi_{\alpha}(x_1,..,x_{N-1},x)|^2, 
\end{eqnarray}which is equivalent to computing
\begin{equation}
n_{B (F)}^\alpha(x)=\sum_{i=1}^N p_{i,B(F)}^\alpha \rho_i(x),
\end{equation}
with $ p^\alpha_{i,B(F)} = 1$ if a boson (fermion) is at position $i=1,...,N$ in the configuration $\alpha$ and zero otherwise, and
\begin{equation}
\rho_i(x)=\int_{x_1<x_2...<x_N}\!\!\!\!\!\!\!\!\!\!\!\!\!\!\!dx_1...dx_N |\Psi_F(x_1,..x_N)|^2 \delta(x-x_i).
\end{equation}

\subsection{An illustration with $N_B=N_F=2$}
We illustrate the idea in the case $N=4$, $N_B=N_F=2$, and take a harmonic
confinement $V(x) = \frac{1}{2} m \omega^2 x^2$ for simplicity, assuming  the same trapping frequency
$\omega$ for the two species. We denote the bosonic coordinates by $x_1$, $x_2$
and the fermionic ones by $x_3$, $x_4$, and we expect a six-fold
degeneracy. We label the basis set using the positions of the particles,
i.e. BBFF, BFBF, BFFB, FBBF, FBFB, FFBB. Let us call this basis the ``BBFF''
basis.  According to the above prescription the first wavefunction is
\begin{equation}\label{eqn:BBFF}
\Psi_{BBFF}=\frac{1}{2}\left[\langle x_1,x_2,x_3,x_4|(e+(12))(e-(34)) \rangle \right],
\end{equation}  
where by $(j\ell)$ we denote the permutation between the particles $j$ and
$\ell$, $e$ is the identity permutation, and we adopt the usual convention of
product among permutations \cite{Hamermesh_book}; the snippet basis satisfies
$\langle x_1,..x_N|P+Q\rangle = \langle x_1,..x_N|P\rangle + \langle
x_1,..x_N|Q\rangle $.  Similarly, the second wavefunction is obtained by
\begin{equation}
\Psi_{BFBF}=\frac{1}{2}\left[\langle x_1,x_3,x_2,x_4|(e+(12))(e-(34)) \rangle \right].
\end{equation}
The other wavefunctions are built in the same way, taking as initial
coordinate sector the one where the bosonic and fermionic coordinates are each
in ascending order.

\begin{figure}
\centering
\includegraphics[width=1.0\linewidth]{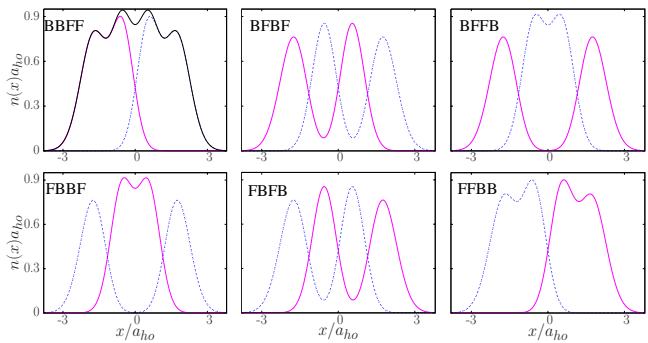}
\caption{\label{fig:rhoBBFF} (Color online) Bosonic (magenta solid line) and
  fermionic (blue dashed line) density profiles (in units of
  $a_{ho}^{-1}=\sqrt{m\omega/\hbar}$) as a function of the spatial coordinate
  $x$ (in units of $a_{ho}$) for each BBFF wavefunction of the $N_B=N_F=2$
  mixture.  The total density (black solid thin line) is included in the first
  frame for reference.  }
\end{figure}

We display the density profiles of the six basis states in
Fig.\ref{fig:rhoBBFF}. Each peak corresponds to the position of a particle in
the BBFF sequence, hence the basis set recalls the one of distinguishable
particles, as in the case of spinor bosons \cite{Deuretzbacher}.

\section{Symmetry characterization} 

\subsection{Casimir invariance of the manifold}
We would like to label the basis vectors by some additional quantum number. We
proceed by exploiting the exchange symmetry between bosons or fermions among
themselves.  According to a group theoretical analysis, there are only two
possible Young tableaus associated to a quantum mechanical system of mixed
Bose-Fermi symmetry, i.e. symmetric in its first $N_B$ coordinates and
antisymmetric in its last $N_F$:
\begin{equation}
\begin{tabular}{ccccc}
$Y$ = &
\begin{tabular}{|c|c|c|c|}
       \hline
       $F_1$ & $B_1$ & $\cdots$ & $B_{N_B}$ \\
       \hline
       \multicolumn{1}{|c|}{$\vdots$} & \multicolumn{3}{|c}{} \\
       \cline{1-1}
       \multicolumn{1}{|c|}{$F_{N_F}$} & \multicolumn{3}{|c}{} \\
       \cline{1-1}
\end{tabular}~,
 & ~ & $Y'$ = &
\begin{tabular}{|c|c|c|}
       \hline
       $B_1$ & $\cdots$ & $B_{N_B}$ \\
       \hline
       \multicolumn{1}{|c|}{$F_1$} & \multicolumn{2}{|c}{} \\
       \cline{1-1}
       \multicolumn{1}{|c|}{$\vdots$} & \multicolumn{2}{|c}{} \\
       \cline{1-1}
       \multicolumn{1}{|c|}{$F_{N_F}$} & \multicolumn{2}{|c}{} \\
       \cline{1-1}
\end{tabular}
\end{tabular}~,
\label{tableau}
\end{equation}
 each with a dimension of $C^{N-1}_{N_B}$,  $C^{N-1}_{N_F}$ respectively. 

To each Young tableau it is possible to associate a value of a Casimir
invariant, obtained from the generators of the permutation group ${\cal S}_N$, and 
which commutes with all elements of the group. In this particular case we
choose ${\cal C}=\sum_{i<j} (ij)$, the sum over all the transpositions
\cite{NovKat04}. Two different
eigenvalues of the Casimir operator are associated to the two
tableaus, namely 
\begin{eqnarray}
 {\cal C}_Y & = & C^{N_B+1}_{2} - C^{N_F-1}_{2},\nonumber\\
{\cal C}_{Y'} & = & C^{N_B-1}_{2} - C^{N_F+1}_{2},
\end{eqnarray}
with degeneracy corresponding to the dimension of the tableau.  We
remark that the eigenvalue ${\cal C}_{Y(Y')}$ essentially counts the number of two-particle
exchange allowed by the corresponding tableau, where a symmetric permutation
(from the row) is counted as $+1$ and an antisymmetric permutation (from the
column) is counted as $-1$.

We represent the
Casimir operator on the BBFF orthonormal basis set (see appendix \ref{appA}
for some examples).  The eigenvectors of the Casimir operators, obtained by
diagonalization, are expressed as linear combinations of the BBFF basis
vectors, and provide an alternative (but not orthonormal) basis set $\Psi_\mu(x_1,...,x_N)$
characterized by the symmetry of the tableau. 

The related bosonic and fermionic momentum distributions
\begin{equation}
n^\mu_{B(F)}(p) = \frac{1}{2\pi} \int dx~dx' e^{-ip(x-x')} \rho^\mu_{B(F)}(x,x')
\end{equation}
can be computed from the bosonic and fermionic one-body density matrix
\begin{eqnarray}
\rho^\mu_{B}(x,x') & = & N_B\!\! \int \!\!dx_2...dx_N
\Psi_\mu^*(x,x_2,...,x_N)\nonumber\\
& & \hspace{2cm}\times \Psi_\mu(x',x_2,...,x_N),\nonumber\\
\rho^\mu_{F}(x,x') & = & N_F\!\! \int \!\!dx_1...dx_{N-1}
\Psi_\mu^*(x_1,..x_{N-1},x)\nonumber\\
& & \hspace{2cm}\times \Psi_\mu(x_1,...x_{N-1},x').  
\end{eqnarray}

\subsection{Density profiles and momentum distributions at a given symmetry}

\begin{figure}
\includegraphics[width=1.0\linewidth]{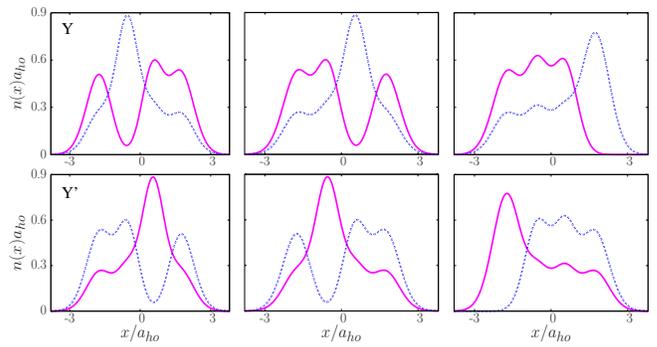}
\caption{\label{fig:rhoCasimir} (Color online) Bosonic (magenta solid line)
  and fermionic (blue dashed line) density profiles (in units of
  $a_{ho}^{-1}$) as a function of the spatial coordinate $x$ (in units of
  $a_{ho}$) with the $Y$ (top) and $Y'$ (bottom) symmetry for $N_B=N_F=2$.}
\end{figure}

We return to the earlier example of two bosons and two fermions in a harmonic
confinement.  The density profiles of the eigenstates of $Y$ and $Y'$ are
shown in Fig.~\ref{fig:rhoCasimir}. We notice that in this highly symmetric
problem the role of bosons and fermions is simply exchanged among the two
submanifolds.

\begin{figure}
  \includegraphics[width=1.0\linewidth]{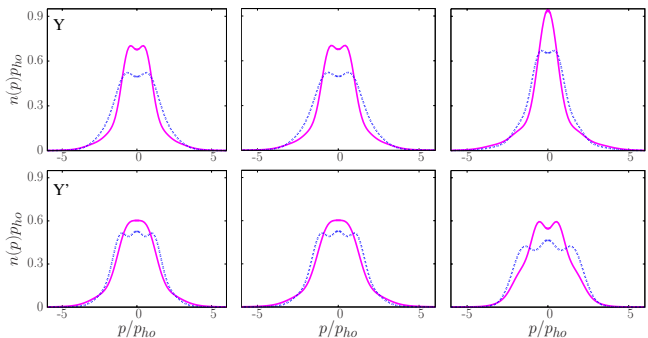}
\caption{\label{fig:momdistrCasimir} (Color online) Bosonic (magenta solid
  line) and fermionic (blue dashed line) momentum distributions (in units of
  $p_{ho}^{-1}=1/\sqrt{\hbar m \omega}$) as a function of $p$ (in units of
  $p_{ho}$) with the $Y$ (top) and $Y'$ (bottom) symmetry for $N_B=N_F=2$. The
  first and second panels in each row coincide.}
\end{figure}

We also display the corresponding momentum distributions in
Fig.\ref{fig:momdistrCasimir}.  It is worth mentioning that the momentum
distributions associated to the $Y$ symmetry display less peaks (with a number of
peaks in $n_F(p)$ equal to $N_F$ as in \cite{Fang09}) and are narrower than
those associated to the $Y'$ symmetry, following the intuition that the
symmetry of the latter tableau is more ``Fermi-like'' (i.e. extended in the
vertical direction).

\section{Ground state at large, finite interaction strength}

\subsection{Analysis of  special solutions}
At finite interactions $g_{BB}=g_{BF}$,
the ground state is expected to display the $Y$ symmetry
\cite{LaiYang,Yang09}, since the associated wavefunction has less nodes than
the one with $Y'$ symmetry. From a continuity argument starting from the
noninteracting solution, we also expect that it is nonvanishing in all
permutation sectors.  In Ref.\cite{GirMin07} a special solution with the
latter property was proposed, 
\begin{equation}
\Psi_{GM}=A_{BB}A_{BF}\Psi_F(x_1,...,x_N),
\end{equation}
where the mapping functions are 
\begin{eqnarray}
A_{BB} & = & \Pi_{1\le j<\ell\le N_B}~{\rm
  sgn}(x_j-x_\ell),\nonumber\\
A_{BF} & = & \Pi_{1\le j\le N_B,N_B+1\le\ell\le N}~{\rm
  sgn}(x_j-x_\ell).
\end{eqnarray}
We test its symmetry by evaluating the average value of
the Casimir operator $\langle {\cal C}\rangle=\langle \Psi| {\cal C} |\Psi\rangle/\langle
\Psi|\Psi\rangle$. The expression of the (unnormalized) GM wavefunction on the
BBFF basis $\Psi_{GM}=\sum_{\alpha=1}^{N!/N_F!/N_B!} c_\alpha \Psi_\alpha $ is
$c_\alpha=1$ for any $\alpha$. From the representation of the Casimir operator
(see appendix \ref{appA}) we obtain that for $N_B=N_F=3$ the average value of
the Casimir operator corresponds to its maximal eigenvalue ${\cal C}_Y=3$, hence the
GM wavefunction has the symmetry of the $Y$ tableau. This is not the case for
$N_B=N_F=2$ where a wavefunction which is nonvanishing in all coordinate
sectors and has the $Y$ symmetry is
$\Psi'_{GM}=A_{BB}\Psi_F(x_1,...,x_N)$. More generally, we have proven that
the GM wavefunction with odd $N_B=N_F$ has always the symmetry of the $Y$
tableau (see appendix \ref{appB} for demonstration).

For the homogeneous system, the Bethe-Ansatz method provides a solution of the
model of a Bose-Fermi mixture with equal bosonic and fermionic masses and
finite but equal coupling strengths $g_{BB}=g_{BF}$
\cite{LaiYang,ImaDem06,ImaDemAnn}. The solution for the many-body wavefunction
is built with the symmetry of a given tableau \cite{Yang67,Sutherland68}. In
analogy with fermionic systems, the solution is expressed in terms of spatial
coordinates and ``pseudospin'' integer coordinates $y_i$ which correspond to
the relative positions of the bosons in the coordinate sector
$x_{P(1)}<x_{P(2)}<...<x_{P(N)}$, namely
$\{y_1,...y_{N_B}\}=\{P^{-1}(1),...,P^{-1}(N_B)\}$.  In Ref.\cite{ImaDemAnn}
Imambekov and Demler study the strongly interacting limit of the Bethe-Ansatz
solution (BA) for $N_B$, $N_F$ odd. They notice that the wavefunction
decouples as a product of an ``orbital'' part and a ``spin'' part, as 
\begin{equation}
\label{psiID}
\Psi_{BA}\sim \det[e^{i \frac{2\pi}{N}\kappa_iy_j}] \Psi_F(x_1,...x_N),
\end{equation}
where the orbital part is a Slater determinant of the first $N$ orbitals, of
course chosen for the homogeneous problem $\Psi_F(x_1,...x_N)=(1/\sqrt{N!})\det[e^{ik_j x_\ell}]$, and the
``spin'' part includes only the bosonic coordinates $y_1,...y_{N_B}$, with
$\kappa=\{-(N_B-1)/2+N/2,...,N/2,...(N_B-1)/2+N/2 \}$ for the ground
state. Taking advantage of the decoupling among orbital and spin part of the
wavefunction, we generalize such a solution to the inhomogeneous system, by
replacing the orbital part of (\ref{psiID}) by its corresponding expression
under confinement, Eq.\ (\ref{eqn:SD}).  
We check its symmetry by evaluating the average of the Casimir operator.  For
$N_B=N_F=3$ an explicit calculation by expansion of (\ref{psiID}) on the BBFF
basis yields the maximal value $\langle \Psi_{BA}| {\cal C} |\Psi_{BA}\rangle/\langle
\Psi_{BA}|\Psi_{BA}\rangle=3$, implying that the generalized BA wavefunction
has the $Y$ symmetry.  By construction, the 
 BA solution for the trapped case  has the same form as the one obtained 
for arbitrary interactions in the homogeneous system \cite{note2}. 
 Notice that although the GM and generalized BA 
 wavefunctions are not
simply proportional to each other, their density profiles coincide, displaying
no demixing.  We remark that
Eq.~(\ref{psiID}) can generate a wavefunction with the $Y'$ symmetry by
adopting a different choice of spin rapidities, e.g. for
$\kappa=\{-(N_B-1)/2,...,0,...(N_B-1)/2\}$ we obtain an average Casimir
operator equal to $-3$.

It is important to notice that the form of the solution for the wavefunction
at finite large interactions depends from the way in which the limit $g_{BB}\to
\infty$ and $g_{BF}\to \infty$ is approached. Consider for example the case
where $g_{BB}$ tends to infinity and $g_{BF}$ is finite. The above symmetry
classifications are not useful in this case. The bosonic component can
be mapped onto a fermionic one, and the problem can be reduced to the one of
spin 1/2 fermions as in \cite{Ogata90}. Also in such a case a decoupling of
spatial and spin degrees of freedom is predicted with a different form for the
spin part with respect to Eq.(\ref{psiID}).

\begin{figure}
\includegraphics[width=0.5\linewidth]{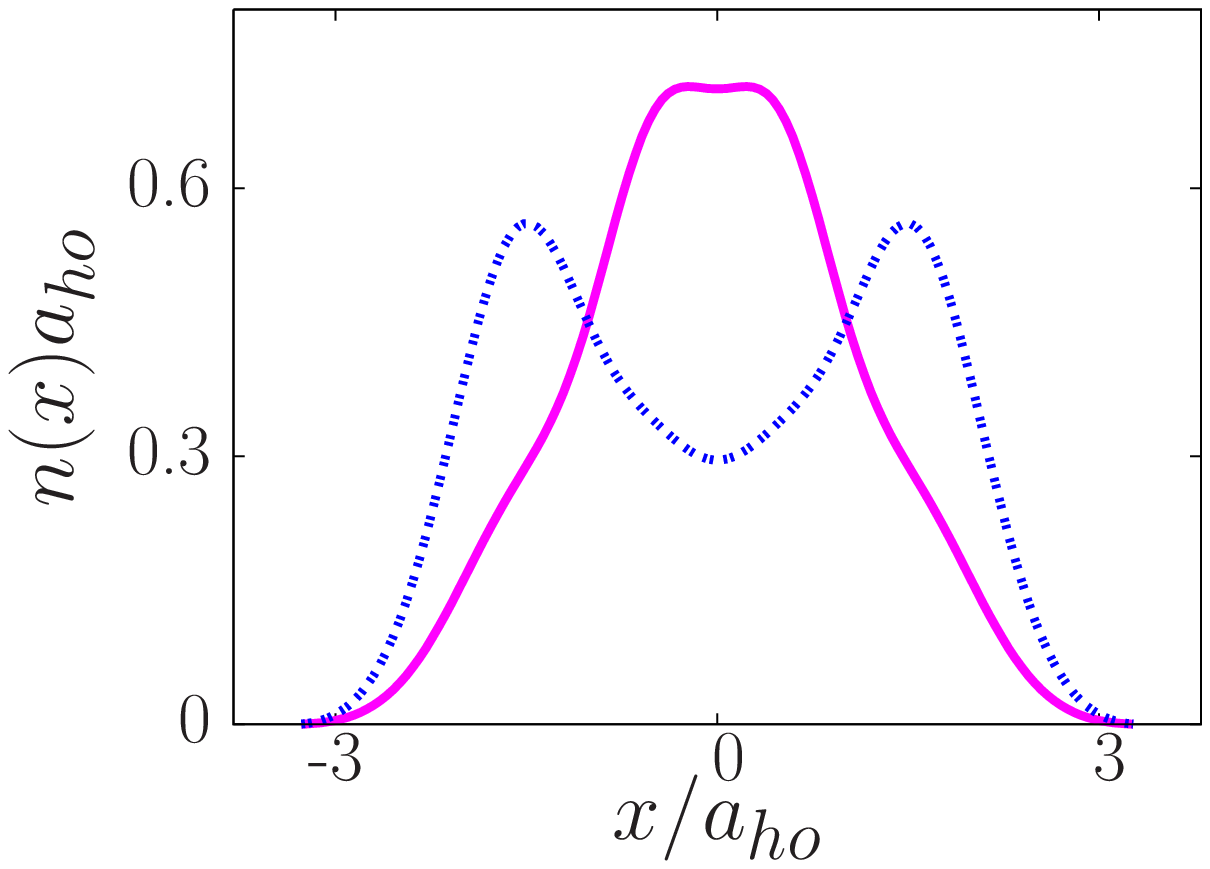}\includegraphics[width=0.5\linewidth]{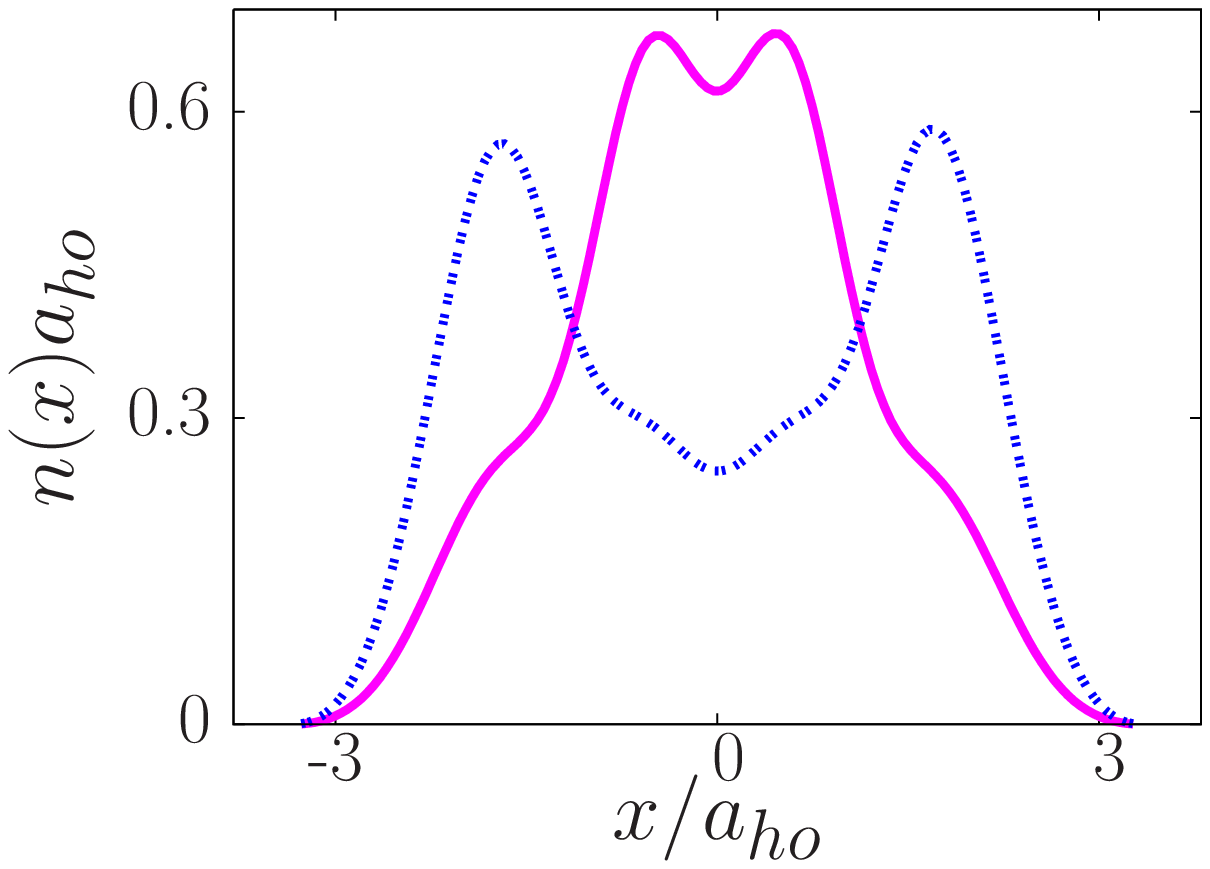}
\includegraphics[width=0.5\linewidth]{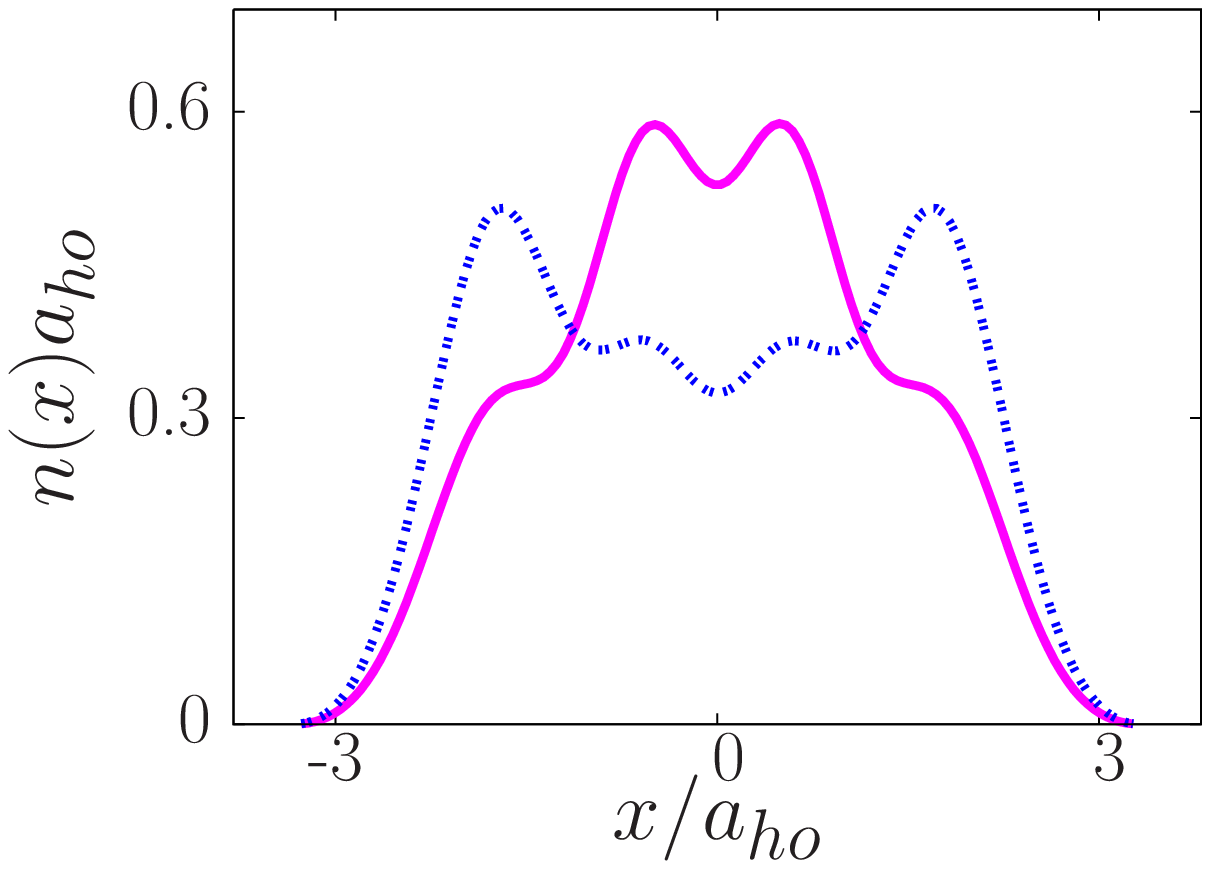}\includegraphics[width=0.5\linewidth]{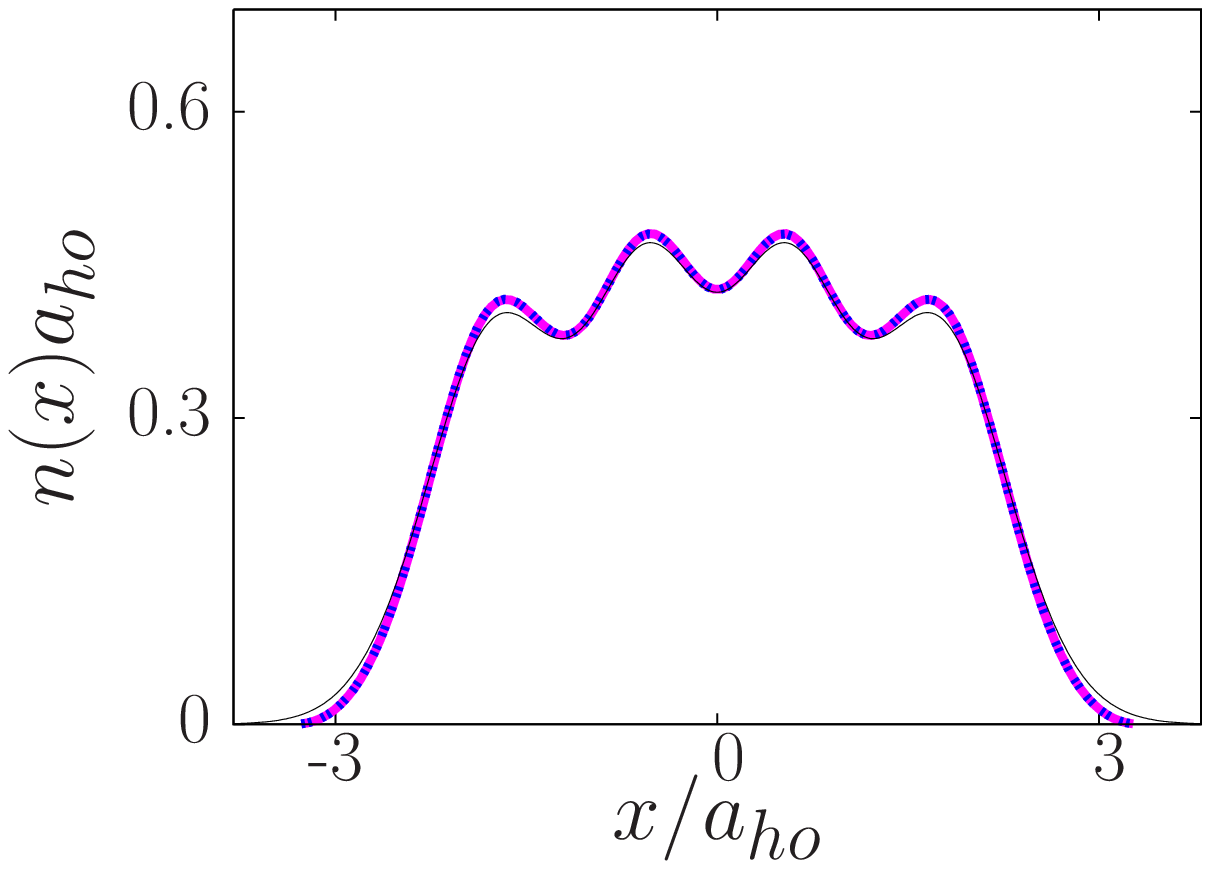}
\caption{\label{fig:densprofDMRG} (Color online) Bosonic (magenta solid
  line) and fermionic (blue dashed line) density profiles as obtained from DMRG simulations (in units of
  $a_{ho}^{-1}$) as a function of $x$ (in units of
  $a_{ho}$) at increasing interaction strength $U/t=1, 10, 10^2, 10^4$ (from left to right and from top to bottom). The harmonic trap strength is $V/t=7 \times 10^{-6}$ and the number of lattice sites used in the simulation is $L=128$. In the last panel the analytical  prediction from the generalized  BA wavefunction (thin black solid line) is shown.}
\end{figure}

\begin{figure}
\includegraphics[width=0.49\linewidth]{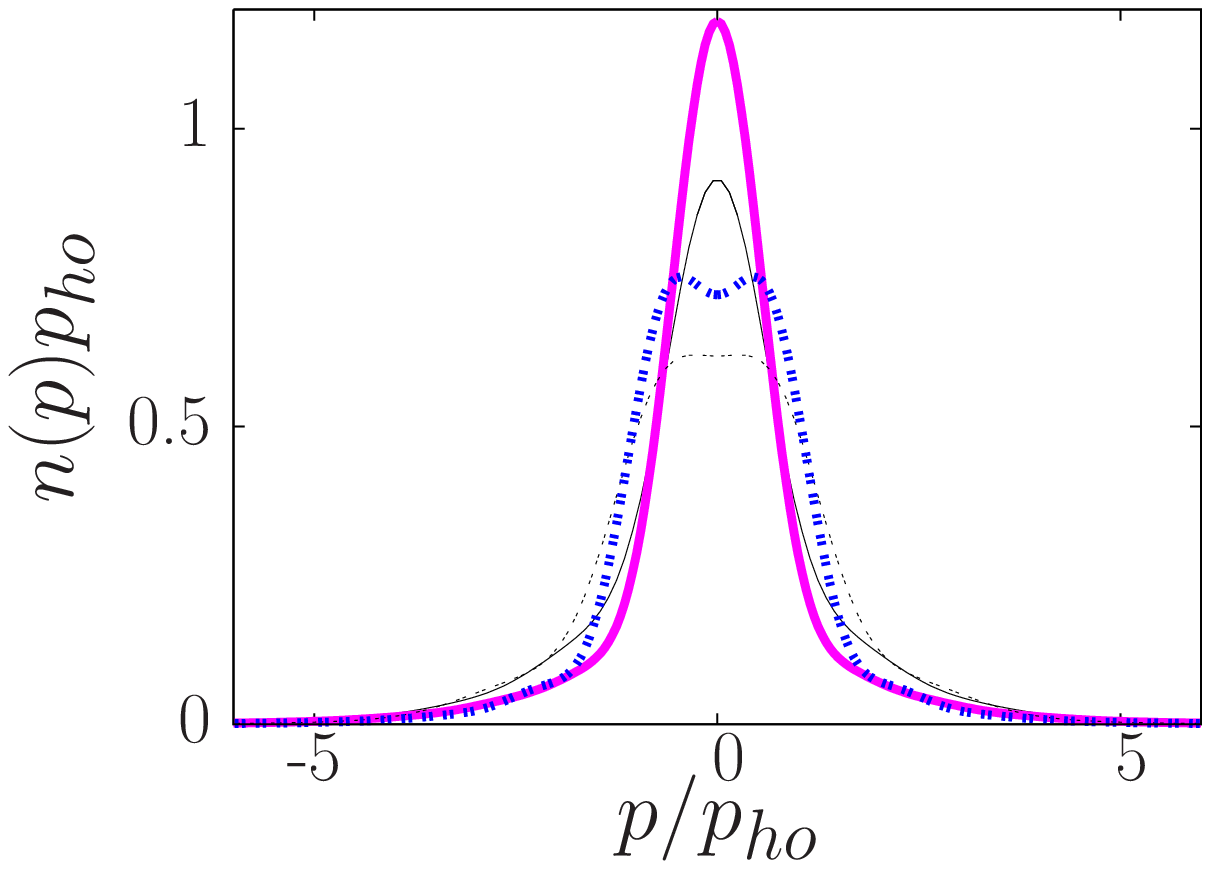}\includegraphics[width=0.49\linewidth]{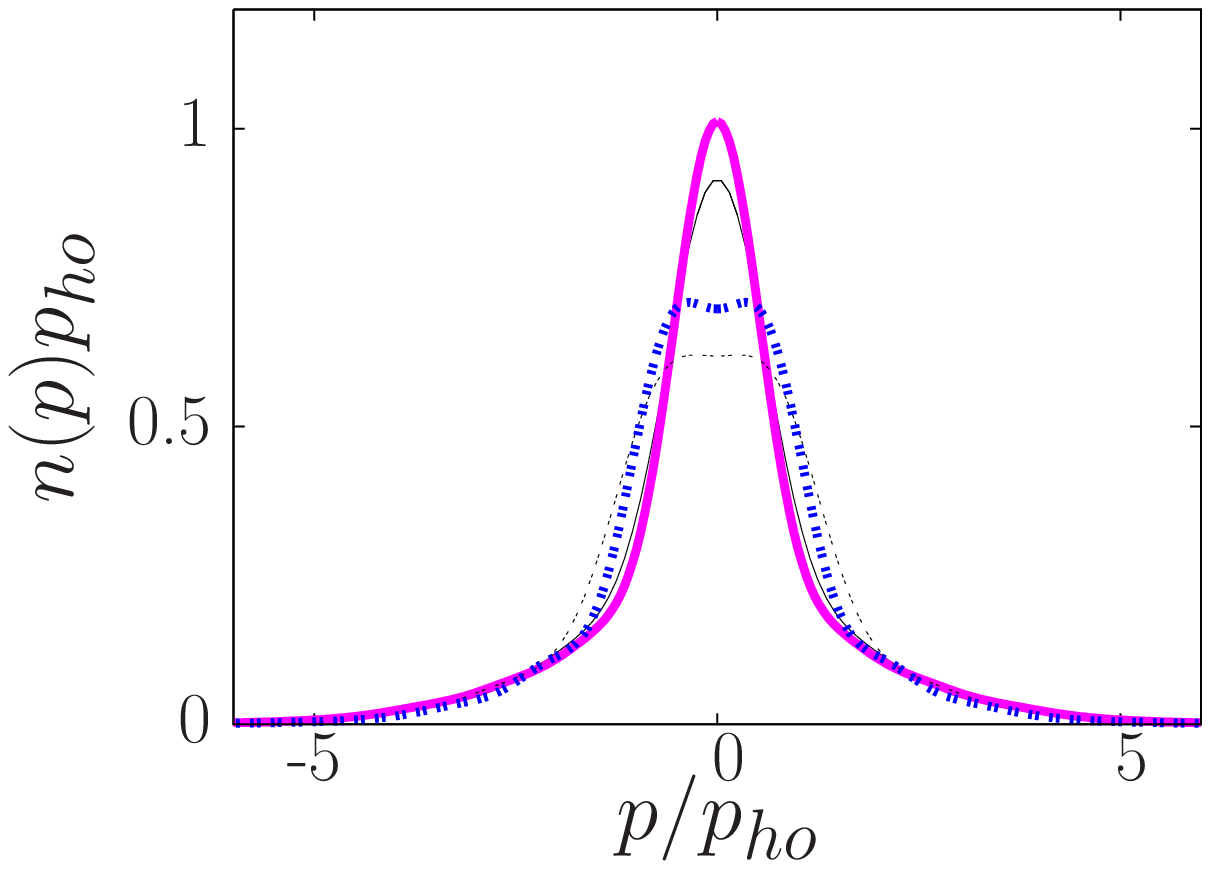}
\includegraphics[width=0.49\linewidth]{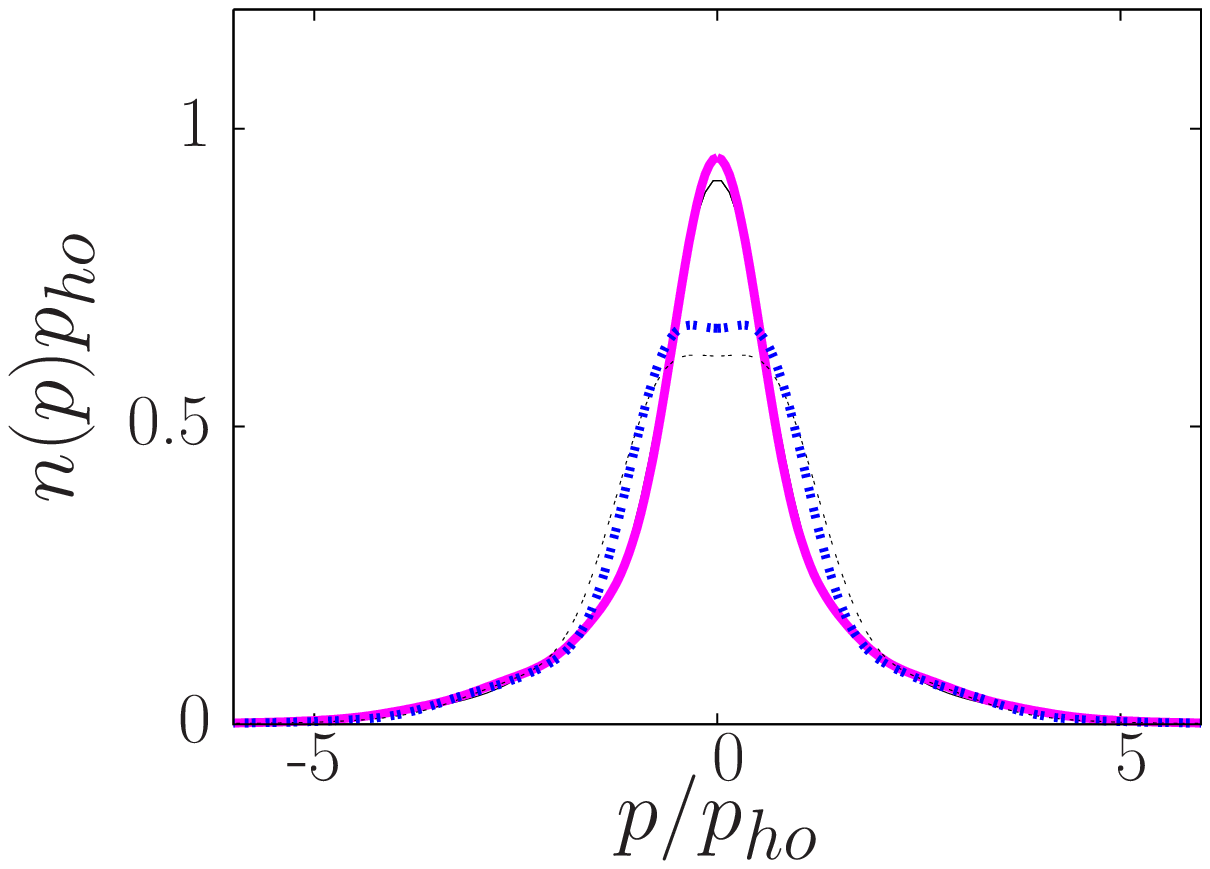}\includegraphics[width=0.49\linewidth]{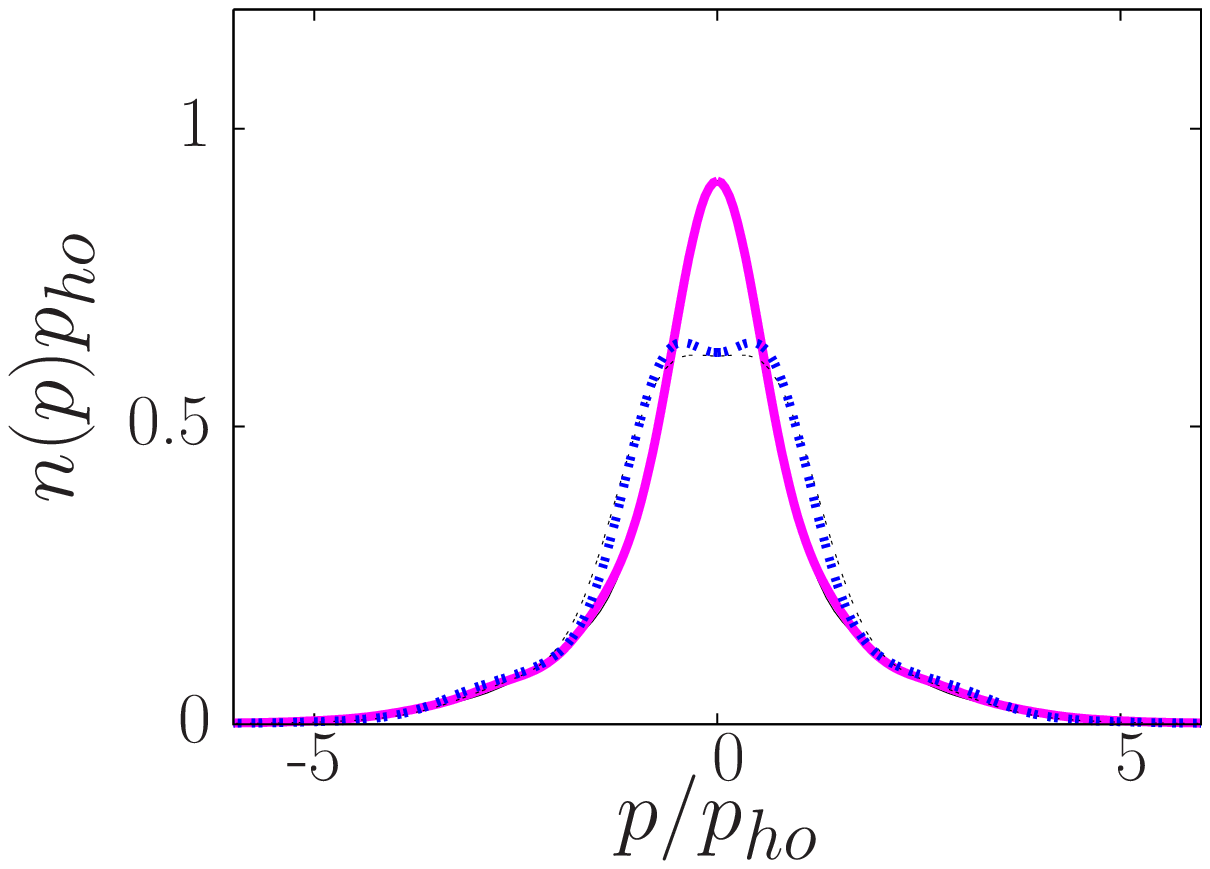}
\caption{\label{fig:momdistrDMRG} (Color online) Bosonic (magenta solid lines) and fermionic (blue dashed lines) momentum distributions  (in units of
  $p_{ho}^{-1}=1/\sqrt{\hbar m \omega}$) as a function of $p$ (in units of
  $p_{ho}$), from the  numerical DMRG data at increasing interaction strength   ($U/t=1, 10, 10^2, 10^4$, from top to bottom and from left to right,  thick lines). At increasing interactions the curves approach the analytical predictions from the generalized BA  wavefunction for the bosonic (thin black solid lines) and fermionic (thin black dashed lines) momentum distributions, shown in each panel for reference (almost indistinguishable from the numerical data in the last panel).  The other parameters used in the simulation are $V/t=7\times 10^{-6}$ and $L=128$.}
\end{figure}

\subsection{Numerical illustration}
We have tested our predictions by comparing with numerical DMRG simulations \cite{linkSNS}. 
The DMRG techniques provides an efficient numerical solution for the lattice model of an interacting Bose-Fermi mixture subjected to an external potential, described by the Hamiltonian 
\begin{eqnarray}
{\cal H}= - t \sum_j (b^\dagger_j b_{j+1}+h.c.) - t \sum_j (f^\dagger_j f_{j+1}+h.c.)\nonumber \\ +  U_{BB} \sum_j n_{B,j} (n_{B,j}-1)+  U_{BF} \sum_j n_{B,j} n_{F,j} \nonumber \\+ \sum_j V (j-L/2)^2 ( n_{B,j}+n_{F,j}).
\end{eqnarray}
Here, $b_j$ and $f_j$ are the bosonic and fermionic field operators acting on a site $j$, with corresponding density operators  $n_{B,j}= b^\dagger_j b_{j}$ and $n_{F,j}= f^\dagger_j f_{j}$,  $t$ is the tunnel constant, $U_{BB}$ and $U_{BF}$ are the on-site BB and BF interactions,  $V$ is the strength of the external harmonic confinement, which is taken  to be the same for bosons and fermions, and $L$ is the number of sites in the lattice.

Since in the limit of very low lattice filling the lattice model reproduces the continuum model, we have used this method to obtain the ground-state density profiles and momentum distributions for the Bose-Fermi mixture at increasing interaction strength. For the sake of comparing with the previously proposed special solutions we have restricted ourselves to  the case of equal BB and BF interactions, $ U\equiv U_{BB}= U_{BF}$. 
Figure \ref{fig:densprofDMRG} displays the bosonic and fermionic density profiles for a Bose-Fermi mixture with $N_B=2$, $N_F=2$ at increasing interaction strength. The density profiles evolves from a partially demixed one at intermediate interaction strength, to a nondemixed one at large interactions, in agreement with the predictions of the GM and generalized BA wavefunctions. This behaviour  was also noticed in a density functional calculation \cite{Hao11}.
 For a Bose-Bose mixture a similar absence of demixing at strong coupling 
was observed \cite{Zollner08}. 

The momentum distributions for the bosonic and fermionic components  are illustrated in Fig.\ref{fig:momdistrDMRG} as compared with the predictions of the generalized BA wavefunction. At increasing the interaction strength the momentum distributions approach those obtained by the generalized BA wavefunction, showing that this solution accurately describes the trapped mixture at finite, large but
equal interactions. 


\section{Outlook and perspectives} 

In this work we have studied the exact solutions of a highly symmetric model
for Bose-Fermi mixture in the strongly interacting limit. Our results are
relevant for the ongoing experiment on ultracold mixtures of atomic gases in
tight atomic waveguides, with particular attention of the case of
$^{173}$Yb-$^{174}$Yb Bose-Fermi mixtures \cite{Fukuhura09} where the
fractional mass difference among the two isotopes is small. We have provided
first an orthogonal basis set of solutions which span the degenerate
manifold. Secondly, we have grouped such solutions on the basis of the Casimir
invariant, associated to a given Young tableau. Finally, we have analyzed two
special solutions of the problem and discussed their symmetry according to the
average value of the Casimir operator. By comparing with DMRG simulations, we have found that the  wavefunction obtained by generalizing the Bethe-Ansatz solution to trapped systems accurately describes the density profile and momentum distribution of the mixture at finite, large but equal BB and BF coupling strengths. This wavefunction can be used to describe the Bose-Fermi mixture in arbitrary external potential.
Splittings and mixing of the states with different symmetries are
expected when different masses and different BB and BF coupling constants are
chosen. Our solution serves as a guideline for further
 numerical studies.  This
work opens also the way to the study of the dynamical properties of the strongly
interacting Bose-Fermi mixture.  Signatures of strong correlations could be
found in the collective excitation spectrum.  It would also be
interesting to investigate how particular states in the degenerate manifold
can be prepared and addressed.

\acknowledgments We thank F. Deuretzbacher for suggestions on
the $N=3$ case, 
M. Rizzi for help with the DMRG code 
and  B. Gr\'emaud for
discussions. This work has been developed by using the DMRG code released within the "Powder with Power" project (www.qti.sns.it). 
 We acknowledge support from the MIDAS STREP project, the Handy-Q ERC project 
and from the CNRS PEPS-PTI ``Quantum gases and condensed matter''. ChM
acknowledges support from the CNRS PICS Grant No. 4159 and from the
France-Singapore Merlion program, FermiCold grant No. 2.01.09.  Centre for
Quantum Technologies is a Research Centre of Excellence funded by the Ministry
of Education and the National Research Foundation of Singapore.

\appendix

\section{Casimir operator on the BBFF basis}
\label{appA}

For $N_B=2$, $N_F=2$ the representation of the Casimir operator ${\cal C}=\sum_{i<j}
(ij)$ on the orthonormal BBFF configuration basis  reads
\begin{equation}
 \left( \begin{array}{cccccc}
0 & 1 & -1 &1&-1 &0 \\
1 & 0 & 1  &1& 0 &-1 \\
-1 & 1 & 0 &0&1 &-1 \\
1 & 1 & 0 &0&1 &1 \\
-1 & 0 & 1 &1&0 &1 \\
0 & -1 & -1 &1&1 &0 \\
\end{array} \right).
\end{equation}

For  $N_B=3$, $N_F=3$ the Casimir operator  reads 
\begin{widetext}
\begin{equation}   
\left(
   \begin{array}{llllllllllllllllllll}
    0 & 1 & -1 & 1 & 1 & -1 & 1 & 0 & 0 & 0 & 1 & -1 & 1 & 0 & 0 & 0 & 0 & 0 &
      0 & 0 \\
    1 & 0 & 1 & -1 & 1 & 0 & 0 & -1 & 1 & 0 & 1 & 0 & 0 & -1 & 1 & 0 & 0 & 0 &
      0 & 0 \\
    -1 & 1 & 0 & 1 & 0 & 1 & 0 & -1 & 0 & 1 & 0 & 1 & 0 & -1 & 0 & 1 & 0 & 0 &
      0 & 0 \\
    1 & -1 & 1 & 0 & 0 & 0 & 1 & 0 & -1 & 1 & 0 & 0 & 1 & 0 & -1 & 1 & 0 & 0 &
      0 & 0 \\
    1 & 1 & 0 & 0 & 0 & 1 & -1 & 1 & -1 & 0 & 1 & 0 & 0 & 0 & 0 & 0 & -1 & 1 &
      0 & 0 \\
    -1 & 0 & 1 & 0 & 1 & 0 & 1 & 1 & 0 & -1 & 0 & 1 & 0 & 0 & 0 & 0 & -1 & 0 &
      1 & 0 \\
    1 & 0 & 0 & 1 & -1 & 1 & 0 & 0 & 1 & -1 & 0 & 0 & 1 & 0 & 0 & 0 & 0 & -1 &
      1 & 0 \\
    0 & -1 & -1 & 0 & 1 & 1 & 0 & 0 & 1 & 1 & 0 & 0 & 0 & 1 & 0 & 0 & -1 & 0 &
      0 & 1 \\
    0 & 1 & 0 & -1 & -1 & 0 & 1 & 1 & 0 & 1 & 0 & 0 & 0 & 0 & 1 & 0 & 0 & -1 &
      0 & 1 \\
    0 & 0 & 1 & 1 & 0 & -1 & -1 & 1 & 1 & 0 & 0 & 0 & 0 & 0 & 0 & 1 & 0 & 0 &
      -1 & 1 \\
    1 & 1 & 0 & 0 & 1 & 0 & 0 & 0 & 0 & 0 & 0 & 1 & -1 & 1 & -1 & 0 & 1 & -1 &
      0 & 0 \\
    -1 & 0 & 1 & 0 & 0 & 1 & 0 & 0 & 0 & 0 & 1 & 0 & 1 & 1 & 0 & -1 & 1 & 0 &
      -1 & 0 \\
    1 & 0 & 0 & 1 & 0 & 0 & 1 & 0 & 0 & 0 & -1 & 1 & 0 & 0 & 1 & -1 & 0 & 1 &
      -1 & 0 \\
    0 & -1 & -1 & 0 & 0 & 0 & 0 & 1 & 0 & 0 & 1 & 1 & 0 & 0 & 1 & 1 & 1 & 0 &
      0 & -1 \\
    0 & 1 & 0 & -1 & 0 & 0 & 0 & 0 & 1 & 0 & -1 & 0 & 1 & 1 & 0 & 1 & 0 & 1 &
      0 & -1 \\
    0 & 0 & 1 & 1 & 0 & 0 & 0 & 0 & 0 & 1 & 0 & -1 & -1 & 1 & 1 & 0 & 0 & 0 &
      1 & -1 \\
    0 & 0 & 0 & 0 & -1 & -1 & 0 & -1 & 0 & 0 & 1 & 1 & 0 & 1 & 0 & 0 & 0 & 1 &
      1 & 1 \\
    0 & 0 & 0 & 0 & 1 & 0 & -1 & 0 & -1 & 0 & -1 & 0 & 1 & 0 & 1 & 0 & 1 & 0 &
      1 & 1 \\
    0 & 0 & 0 & 0 & 0 & 1 & 1 & 0 & 0 & -1 & 0 & -1 & -1 & 0 & 0 & 1 & 1 & 1 &
      0 & 1 \\
    0 & 0 & 0 & 0 & 0 & 0 & 0 & 1 & 1 & 1 & 0 & 0 & 0 & -1 & -1 & -1 & 1 & 1 &
      1 & 0
   \end{array}
   \right).
\end{equation} 
\end{widetext}

\section{Demonstration of the symmetry of a special wavefunction}
\label{appB}

We prove here below that for odd $N_B=N_F$ the GM wavefunction has the
symmetry of the $Y$ tableau.  This readily follows from the simple expression
of the GM wavefunction on the BBFF basis
$\Psi_{GM}=\sum_{\alpha=1}^{N!/N_F!/N_B!} \Psi_\alpha $, and from the fact
that for odd values of $N_F$ and $N_B$ each line of the matrix which
represents the Casimir operator on the BBFF basis has $N_B N_F$ nonvanishing
entries with value 1 in $N_B( N_F+1)/2$ cases and value -1 in $N_B( N_F-1)/2$
cases. Each line of the matrix corresponds then equally to the value of the
average of the Casimir operator, $\langle {\cal C}\rangle=\langle \Psi|{\cal C}
|\Psi\rangle/\langle \Psi|\Psi\rangle$, and its value corresponds to the
maximal eigenvalue ${\cal C}_Y=N_B=N_F$.  For other values of $N_B$, $N_F$ the
representation of the Casimir operator is more complicated to predict and the
above demonstration does not hold.


\begin{thebibliography}{10}
\bibitem{note2}
Note that in the homogeneous limit the GM wavefunction does not coincide with
the Bethe-Ansatz solution at strong coupling.
\end{thebibliography}
\end{document}